# Atom ionization by ultrashort laser pulse – 2D, 3D consideration

N V Larionov[1], D N Makarov[2] and A A Smirnovsky[3]

[1] Peter the Great St. Petersburg Polytechnic University, St. Petersburg, 195251, Russia
[2] Northern (Arctic) Federal University, Arkhangelsk, 163002, Russia
[3] Ioffe Physical-Technical Institute of the Russian Academy of Sciences,
St. Petersburg, 194021, Russia

E-mail: larionov.nickolay@gmail.com

**Abstract.** We theoretically analyze the quantum vortices – the zeros of electron wave function which are formed through atom ionization by ultrashort laser pulse. In 2D space we consider the case of above-threshold ionization of hydrogen atom and in 3D space the ionization of the atom in the zero-radius-potential approximation. With the help of obtained analytical results we make the assumption that localization of quantum vortices in 3D space can be predicted by analysis of a 2D model.

## 1. Introduction

In our previous works [1, 2], using the numerical method of solving the time-dependent Schrödinger equation in expanding space [3], we investigated appearance and evolution of quantum vortices [4, 5] formed upon above-threshold ionization of a two-dimensional (2D) hydrogen atom by an ultrashort laser pulse. These vortices were identified as forbidden regions for the unbound electron in ordinary and momentum space. The identification also included the study of probability current. To clarify the nature of these vortices, the problem was analyzed with the help of time-dependent perturbation theory and it was shown that the vortices arise due to the interference of electron states related to continuous spectrum. These states, in turn, are formed by a two-photon transition through intermediate states of the continuous spectrum [2].

Numerical simulation of the three-dimensional (3D) ionization process poses considerable difficulties. Therefore, it is natural to try to predict the features of the ionization process in the 3D space (in particular, the vortex formation) using the solutions in a space with a lower dimension.

In this paper, ionization of 3D quantum system, which is an electron bound initially by a zero-radius-potential (ZRP), will be analyzed by means of time-dependent perturbation theory. The obtained results will be compared with the results from a problem of 2D-hydrogen atom ionization.

In addition, the ZRP is a good model for various systems and has an independent interest not associated with an atom [6-8]. Nowadays, there is also interest to lower-dimensional systems, in particular to the 2D-hydrogen atom model [9-11], because of various specific phenomena arising there.

## 2. Theoretical description of the atom ionization

The Hamiltonian for atom interacting with the classical electromagnetic field is given by

$$\hat{H} = -\frac{1}{2}\nabla^2 + U(r) - \hat{\vec{d}}\vec{E}(t), \qquad (1)$$

where the first two terms relate to unperturbed atom, $\hat{\vec{d}}$ is the atom dipole moment operator and $\vec{E}(t)$ is the vector of classical electric field. In this paper we modulate ultrashort laser pulse as a linearly polarized field $\vec{E}(t) = \vec{e}_x \tilde{E}(t)$ with the following time dependence of amplitude

$$\tilde{E}(t) = \tilde{E}_0 \cos(\omega t)\left[\theta(T-t) - \theta(-t)\right], \qquad (2)$$

where $\omega$ is the frequency, $T$ is a pulse duration, $\tilde{E}_0$ is a constant amplitude and $\theta$ is the Heaviside step function.
Here and further we use Hartree atomic units: $e = \hbar = m_e = 1$.

*2.1. 2D-hydrogen atom ionization*
In our previews paper [2] we applied time-dependent perturbation theory to 2D-hydrogen atom ionization problem and obtained the probability amplitude that unbound electron have arbitrary projection of the momentum on one axis with fixed zero projection on other axis. Here we generalize our previous results and present expression for the case of arbitrary orientation of electron momentum $\vec{k} = (k_x, k_y) = (k\cos(\varphi_k), k\sin(\varphi_k))$

$$\begin{aligned} b(k, \varphi_k, t) = &-i \sum_{m=\pm 1} b^{(1)}_{km,10}(t)\Phi_m(\varphi_k) \\ &+ b^{(2)}_{k0,10}(t)\Phi_0(\varphi_k) - \sum_{m=\pm 2} b^{(2)}_{km,10}(t)\Phi_m(\varphi_k), \end{aligned} \qquad (3)$$

where $k, \varphi_k$ are polar components of the electron momentum and $\Phi_m(\varphi_k) = e^{im\varphi_k}/\sqrt{2\pi}$. The amplitudes $b^{(1,2)}_{km,10}(t)$ define the expansion of the electron wave function in a set of cylindrical waves. The lower indexes $km,10$ indicate the electron transition from the initial atomic ground state with quantum numbers $n = 1$, $m = 0$ and energy $E_1 = -0.5$ [9, 10] to continuum spectrum characterized by angular momentum projection $m = 0, \pm 1, \pm 2, \ldots$ on the axis z and energy $E_k = k^2/2 = (k_x^2 + k_y^2)/2$. The exact expressions for $b^{(1,2)}_{km,10}(t)$ are presented in [2]. The upper indexes "(1)" and "(2)" correspond to first and second orders of applied perturbation theory respectively.
The amplitude (3), as well as the same amplitude in [2], was found in the approximation of one level atom and by replacement of Coulomb waves by cylindrical waves. One of the reasons of such approximation is that the above-threshold ionization was considered.
Let us remind, that for given laser pulse (2) we observed quantum vortices as zeros of amplitude $b(k_x = 0, k_y, T)$ [2]. These zeros were interpreted as interference of unbound electron states which were result of above-threshold two-photon ionization through intermediate states of continuous spectrum.

*2.2. ZRP atom ionization*
In the 3D space we consider the electron bound initially by ZRP: $U(r) = -\alpha\delta(\vec{r})$. When $\alpha = 1$ then the energy of a single discrete state is equal to the ground state energy of 2D-hydrogen atom. The exact wave functions related to continuous spectrum of electron in a field of ZRP are spherical waves. These facts led us to the idea to compare results from above 2D-hydrogen atom ionization problem with the results obtained from a model of ZRP atom ionization in 3D space.

To obtain the probability amplitude that the electron initially bound by ZRP, after interaction with the laser pulse (2) becomes unbound with momentum $\vec{k} = (k_x, k_y, k_z)$ we make the same manipulations as in [2]. The result is

$$b(k, \theta_k, \varphi_k, t) = -i \sum_{m'=\pm 1} \frac{1}{\sqrt{k}} b^{(1)}_{k1m',100}(t) Y_{1m'}(\theta_k, \varphi_k)$$
$$+ \frac{1}{\sqrt{k}} b^{(2)}_{k00,100}(t) Y_{00}(\theta_k, \varphi_k) - \sum_{m'=0,\pm 2} \frac{1}{\sqrt{k}} b^{(2)}_{k2m',100}(t) Y_{2m'}(\theta_k, \varphi_k), \quad (4)$$

where $k, \theta_k, \varphi_k$ are spherical components of the electron momentum and $Y_{lm}(\theta_k, \varphi_k)$ is a spherical function. The amplitudes $b^{(1,2)}_{klm,100}(t)$ define the expansion of the electron wave function in a set of spherical waves. The lower indexes $klm,100$ indicate the electron transition from the initial ZRP atom single ground state with quantum numbers $n=1$, $l=0$, $m=0$ and energy $E_1 = -0.5$ to continuum spectrum characterized by angular momentum $l = 0,1,2,...$, magnetic quantum number $m = 0,\pm 1, \pm 2,...\pm l$ and energy $E_k = k^2/2 = (k_x^2 + k_y^2 + k_z^2)/2$. The amplitudes $b^{(1,2)}_{klm,100}(t)$ are

$$\begin{cases} b^{(1)}_{k11,100}(t) = \frac{i}{\sqrt{6}} r_{k1,10} \int_0^t e^{i\omega_{k1}t'} \tilde{E}(t') dt', \\ b^{(2)}_{k00,100}(t) = \frac{2i}{\sqrt{6}} \left( \frac{3}{2k} + \frac{d}{dk} \right) \int_0^t b^{(1)}_{k11,100}(t') \tilde{E}(t') dt', \\ b^{(2)}_{k20,100}(t) = -\frac{2i}{\sqrt{30}} \left( \frac{3}{2k} - \frac{d}{dk} \right) \int_0^t b^{(1)}_{k11,100}(t') \tilde{E}(t') dt', \\ b^{(2)}_{k22,100}(t) = \frac{i}{\sqrt{5}} \left( \frac{3}{2k} - \frac{d}{dk} \right) \int_0^t b^{(1)}_{k11,100}(t') \tilde{E}(t') dt', \end{cases} \quad (5)$$

where $\omega_{k1} = E_k - E_1$ is transition frequency and matrix element $r_{k1,10} = 4\sqrt{\frac{1}{\pi}} \frac{k^{3/2}}{(k^2+1)^2}$. In (5) we taken into account the following relation between amplitudes: $b^{(1)}_{k11,100}(t) = -b^{(1)}_{k1-1,100}(t)$, $b^{(2)}_{k22,100}(t) = b^{(2)}_{k2-2,100}(t)$.

## 3. Results

We take the following parameters of the laser pulse $\omega = \pi$, $T = 4$, $\tilde{E}_0 = 0.5$. For these parameters the vortices were identified in 2D case [2].

In the Figure 1a) using the obtained expressions (3), (4) we plot the dependence of probabilities (for convenient, we plot $|b(\vec{k},t)|$ normalized on right resonance maximum value) on the momentum projection $k_x$, while other momentum projections are fixed: $|b(k_x,0)| \equiv |b(k_x, k_y = 0, t > T)|$ - for

2D case; $|b(k_x,0,0)| \equiv |b(k_x, k_y=0, k_z=0, t>T)|$ - for 3D case. One can see that $k_x$- dependence of probabilities in 2D case and in 3D case are very similar to each other. In both cases the highest peaks of probabilities correspond to resonance values of momentum $k_x = \pm\sqrt{2(\omega+E_1)}$ and the asymmetry of the curves is due to finite duration of the laser pulse.

The $k_y$-dependence of the probability $|b(0,k_y)|$ practically coincides with the corresponding dependence of $|b(0,k_y,0)|$ (Figure 1b)). In 2D case, the zeros $|b(0,k_y \approx \pm 2.3)|=0$, where $|b(k_x \approx \pm 2.3, 0)| \approx 1$ has maximum values, are the manifestation of the vortices [2].

From the symmetry of considered problem it is natural to assume that vortices in 3D space can be identified as zeroes of the wave function in $(k_y, k_z)$ plane. This assumption is partially confirmed in Figure 2, where probability $|b(k_x, k_y, k_z)|$ calculated in 3D momentum space (for better resolution of forbidden area for electron, we present density plot for $\ln(|b(k_x, k_y, k_z, t>T)|)$ in the condition $k_x^2 + k_y^2 + k_z^2 \leq 9$, $-3 \leq k_z \leq 0$). One can see that zeros of $|b(k_x, k_y, k_z)|$ are the «circle» lying in the plane $(k_y, k_z)$ which is perpendicular to applied force.

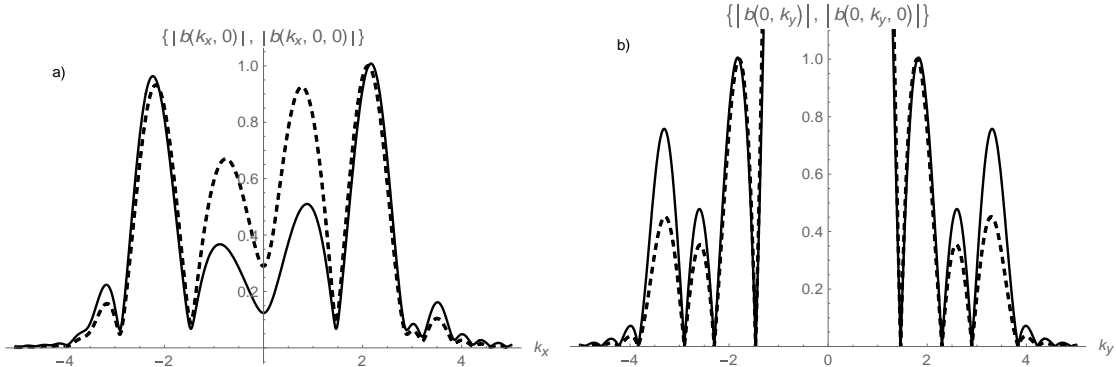

**Figure 1.** Normalized probabilities for 2D case (dashed line, Eq.(3)) and 3D case (solid line, Eq.(4)). a) $|b(k_x,0)|$ and $|b(k_x,0,0)|$, b) $|b(0,k_y)|$ and $|b(0,k_y,0)|$. The distributions are normalized to their right maximum values. The laser pulse parameters: $\omega = \pi$, $T=4$, $\tilde{E}_0 = 0.5$.

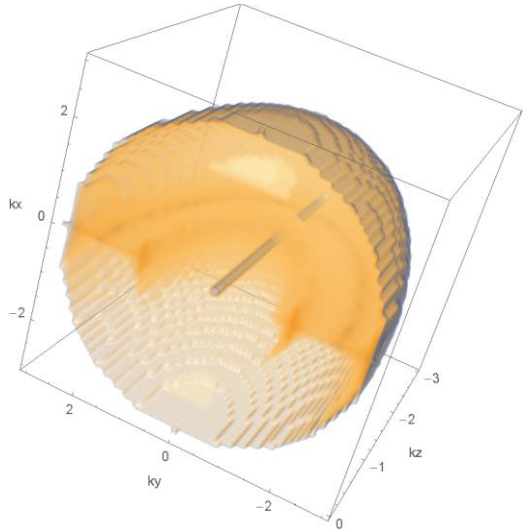

**Figure 2.** The density plot in 3D momentum space for $\ln\left(\left|b\left(k_x,k_y,k_z,t>T\right)\right|\right)$ in the conditions $k_x^2 + k_y^2 + k_z^2 \leq 9$, $-3 \leq k_z \leq 0$. The laser pulse parameters are the same as in Figure 1.

## 4. Conclusion
In this paper we tried to identify the localization of the quantum vortices in three dimensional space by analysis of the problem with lower dimension. We previously reported about vortices identification in the model of 2D-hydrogen atom ionization by ultrashort laser pulse [2]. As a first step to the theory of detection of vortices in 3D space we considered zero-radius-potential atom affected by the laser pulse. The comparison of the results from these two problems with different dimensions allow us to identified the zeroes of wave function in 3D case which can be probably associated with center of quantum vortices.

**Acknowledgments**
The authors gratefully acknowledge the informative discussions with Professor S. Yu. Ovchinnikov from Ioffe Physical-Technical Institute of the Russian Academy of Sciences.